\documentclass[twocolumn,showpacs,preprintnumbers,amsmath,amssymb, prb]{revtex4}

\usepackage{graphicx}
\usepackage{dcolumn}
\usepackage{bm}

\begin{document}


\title{Correlated cluster mean-field theory for spin systems}

\author{Daisuke Yamamoto}
\email{yamamoto@kh.phys.waseda.ac.jp}
\affiliation{Department of Physics, Waseda University, Okubo, Shinjuku-ku, Tokyo 169-8555, Japan }
\homepage{http://www.kh.phys.waseda.ac.jp/}
\date{\today}

\begin{abstract}
A cluster mean-field method is introduced and the applications to the Ising and Heisenberg models are demonstrated. We divide the lattice sites into clusters whose size and shape are selected so that the equivalence of all sites in a cluster is preserved. Since the strength of interactions of a cluster with its surrounding clusters is strongly dependent on the spin configuration of the central cluster itself, we include this contribution in the effective fields acting on the spins. The effects of ``correlations'' between clusters can be taken into account beyond the standard mean-field level, and as a result our cluster-based method gives qualitatively (and even quantitatively) correct results for the both Ising and Heisenberg models. Especially, for the Ising model on the honeycomb and square lattices, the calculated results of the critical temperature are very close (overestimated by only less than $5$ \%) to the exact values.

\end{abstract}

\pacs{75.10.-b, 05.50.+q}
\maketitle
\section{\label{1}Introduction} 
In general, it is difficult to find the exact solution of an interacting many-body system, except for a few simple cases. Hence, various mean-field theories~\cite{PWeiss,Bethe,ORF,Kikuchi,Oguchi} have been proposed and widely used to get some insight into the properties and behavior of the systems. A mean-field concept was first used by Weiss~\cite{PWeiss} for the problem of ferromagnetism. Weiss's molecular-field theory (MFT) was successful in providing a qualitative understanding of the transition from ferromagnetic to paramagnetic states. After the initial success by Weiss,\cite{PWeiss} the MFT has played an important role in the studies of many-body systems and is still often used as a starting point of the investigation due to its simplicity. 

In the vicinity of the critical point, mean-field-type theories generally breaks down since statistical fluctuations neglected in them become important. Therefore, in order to study the detailed critical properties (e.g., critical exponents), it is necessary to resort to other techniques, such as the renormalization-group methods.\cite{Wilson,Fisher,Kadanoff} However, since mean-field-type methods are capable of predicting the behavior of the system over a wide range of parameters (e.g., temperature) at a relatively low cost, a further development of those methods is still important, and many attempts have been made even in recent years.\cite{Wysin,Wysin2,Du,Etxebarria,Zhuravlev} 

The Ising model is the simplest nontrivial model of cooperative phenomena. The Hamiltonian for this model is given by
\begin{eqnarray}
H=-J\sum_{\langle i,j\rangle}\sigma_i \sigma_j, \label{Ising}
\end{eqnarray}
where the sum runs over all nearest-neighbor pairs, $J$ is the coupling strength, and the spin $\sigma_i$ takes either $+1$ or $-1$. In the MFT, the interactions between a spin and its surrounding spins are treated approximately, and the many-body problem is reduced to an effective one-body problem of a single spin in the effective magnetic field $h_{\rm eff}=zJm$, where $z$ is the number of nearest neighbors and $m$ is the magnetization of the system. Solving the self-consistent equation for the magnetization $m=\langle \sigma \rangle$, where $\langle \cdots \rangle$ is an average over the ensemble, one can obtain the well-known result for the critical temperature of the transition from ferromagnetic to paramagnetic states: 
\begin{eqnarray}
k_{B}T_c/J=z. \label{TcMFT}
\end{eqnarray}
Mean-field-type methods generally give only classical predictions for the critical exponents, but the value of the calculated critical temperature is a reasonable measure for assessing the accuracy of the method. 

To obtain more accurate results than this ``one-site cluster'' approximation, it is quite natural that one attempts to increase the size of clusters.~\cite{Bethe,Kikuchi,Oguchi,Etxebarria,Peierls,Weiss,Suzuki,Neto} For example, in the so-called Bethe-Peierls-Weiss (BPW) approximation,\cite{Bethe,Peierls,Weiss} using a cluster of $(z+1)$ sites (one central spin and its surrounding $z$ spins), one can partially take into account the effects of spin correlations and fluctuations. The interactions of the central spin with its nearest neighbors (called the ``first shell'') are treated exactly, while the influence of spins outside the cluster is replaced by an effective field $h_{\rm eff}$ which acts on the $z$ spins of the first shell. Unlike the MFT, the effective field $h_{\rm eff}$ is determined by the condition that the average value of the central spin should be equal to that of the spin on a first-shell site. Recently, Du $et$ $al$.\cite{Du} extended the BPW method by using a group of chains composed of a central chain and its nearest-neighbor chains instead of the cluster of $(z+1)$ sites. Also, Etxebarria $et$ $al$.\cite{Etxebarria} proposed another extended BPW method and demonstrated that it yields a fairly accurate estimate of critical temperature for the square-lattice Ising model. 

Oguchi's method~\cite{Oguchi,Dantziger} is a more straightforward way to improve the results of the MFT. In this method, one considers a cluster consisting of $N_c$ neighboring spins. The interactions between the spins at the cluster edge and the outside spins are treated approximately as effective internal fields. Calculating the average $\langle \sigma_i \rangle$ for each site in the cluster, one regards the mean value of them as the magnetization of the system, i.e., $m=(1/N_c)\sum \langle \sigma_i \rangle$. Then the effective fields are determined self-consistently by the value of $m$. Clearly, the case of $N_c=1$ corresponds to the conventional MFT. Although it is expected that one can obtain closer results to the exact values as the size of the cluster $N_c$ is increased, this method has two shortcomings. First, one has to deal with a quite large cluster to get sufficiently accurate results. In practice, it is difficult to solve exactly the $N_c$-site problem of such a large cluster. Second, since the system is divided into finite-size clusters, equivalence of all sites (or periodicity of the system) is generally lost with few exceptions. 

Recently, Wysin and Kaplan\cite{Wysin} made a significant improvement to the MFT in a quite simple way. Their ``self-consistent'' correlated field (SCCF) approximation remains basically the same as the standard MFT in the sense that one deals with effective one-site problems. In this method, the effective field acting on a central spin from its surrounding spins takes two different values $h_{\rm eff}^{\pm}=zJm^{\pm}$, depending on the state of the spin itself ($\sigma_i=+1$ or $-1$). Here $m^{+}$ ($m^-$) is the average value of the spin on a first-shell site when the state of the central spin is fixed to be $+1$ ($-1$). Taking into account the effects of correlations between neighboring spins in this way, they obtained more accurate results of critical points compared with some other methods, such as the BPW approximation and the Onsager reaction field (ORF) correction.\cite{ORF,Wysin2,White} 

In this paper, as an attempt to further improve these methods, we introduce a new cluster-based approximation method, which we refer to as the ``correlated cluster mean-field'' (CCMF) theory. Depending on the lattice type, we select a cluster of a different size and shape. For example, the calculation for a square lattice model is based on square-shaped four-site ($2\times2$) clusters. Clearly, the four sites in this cluster are equivalent by symmetry. Additionally, to take into account the effects of the cluster-cluster correlations beyond the standard mean-field level, we use a similar idea as in the SCCF method of Ref.~\onlinecite{Wysin}. 

The paper is organized as follows. First, in Sec.~\ref{2}, we demonstrate the application of our modified cluster mean-field theory to the Ising model on several lattices. In Sec.~\ref{3}, the accuracy of our method is verified by comparing the obtained results with those obtained by some other methods. In Sec.~\ref{4}, the cluster-size dependence of the results is discussed. Then, in Sec.~\ref{5}, taking the isotropic Heisenberg model in a uniform field as an example, we extend our approach to quantum spin systems. Finally, a summary is presented in Sec.~\ref{6}. 

\section{\label{2}Correlated cluster mean-field theory for the Ising model}
In the following, we shall demonstrate the application of our method to the Ising model on typical four types of lattices (honeycomb, triangular, square, and simple cubic) one by one. 
\subsection{\label{2-1}Honeycomb lattice}
\begin{figure}[t]
\begin{center}
\includegraphics[width=88mm]{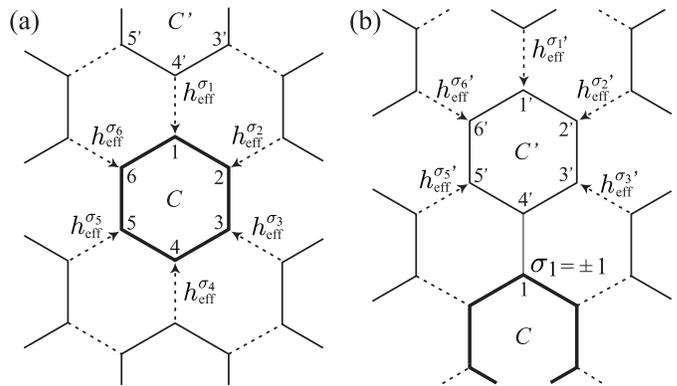}
\caption{\label{fig1} Schematic representation of the honeycomb lattice divided into hexagonal clusters. The arrows indicate the effective fields acting on the sites (a) in cluster $C$ and (b) in cluster $C'$ when the spin at site $1$ is fixed to be $+1$ or $-1$. }
\end{center}
\end{figure}

When the coordination number $z$ is $3$, unfortunately, the SCCF approximation yields a wrong result: the equation for $T_c$ has no solution.\cite{Wysin} It is therefore important to verify whether the critical point can be calculated by applying our method to the case of the two-dimensional (2D) honeycomb lattice. We first divide the lattice into six-site hexagonal clusters as seen in Fig.~\ref{fig1}(a). The strength of interactions of a cluster with its surrounding clusters should be strongly dependent on the spin configuration of the central cluster itself. Thus, including this contribution into the effective fields acting on the spins, we consider the following six-site problem on cluster $C$: 
\begin{eqnarray}
H_{C}=-J\!\!\sum_{\langle i,j\rangle \in C}\!\!\sigma_i \sigma_j-\sum_{i \in C}h_{\rm eff}^{\sigma_i}\sigma_i, \label{Honeycomb}
\end{eqnarray}
where the first (second) sum runs over all nearest-neighbor pairs (all sites) within cluster $C$, and $h_{\rm eff}^{\sigma_i}=Jm^{\sigma_i}$ is the effective field acting on site $i$ from the neighboring spin in the nearby connected cluster (namely, e.g., from the spin at site $4'$ for $i=1$). Here $m^{\sigma_i}$ is the {\it mean field} of the neighboring spin of site $i$ and we assume that its value depends on the state of spin $i$, 
\begin{eqnarray}
m^{\sigma_i}=\left\{ \begin{array}{ll}
m^+ & (\sigma_i=+1) \\
m^- & (\sigma_i=-1). \label{MF_Honeycomb}
\end{array} \right.
\end{eqnarray}
Especially in this case, since there is only one bond between two clusters, the situation is similar to that in the SCCF approximation. Note that the values of $m^+$ and $m^-$ do not depend on the site number $i$ due to the symmetry of the hexagonal cluster. 

When the values of $m^+$ and $m^-$ are obtained, the six-site problem described in Eq.~(\ref{Honeycomb}) is exactly solvable, and we can calculate the magnetization given by
\begin{eqnarray}
m=\langle \sigma_i\rangle={\rm Tr}\left(\sigma_i e^{-\beta H_C}\right)/{\rm Tr}\left(e^{-\beta H_C}\right),~~(i\in C), \label{m}
\end{eqnarray}
where $\beta=1/k_{ B}T$. 

The values of $m^+$ and $m^-$ are determined in the following self-consistent way. Let us now focus on the cluster next to $C$, which we call cluster $C'$ [see Fig.~\ref{fig1}(b)]. One obtains the values of $m^+$ and $m^-$ by calculating the average values of the spin at, for example, site $4'$ in Fig.~\ref{fig1}(b) when the spin at site $1$ is fixed to be $+1$ and $-1$, respectively, i.e., 
\begin{eqnarray}
m^\pm=\langle \sigma_{4'}\rangle \big|_{\sigma_1=\pm 1}={\rm Tr}\left(\sigma_{4'} e^{-\beta H^\pm_{C'}}\right)/{\rm Tr}\left(e^{-\beta H^\pm_{C'}}\right), \label{mpm_Honeycomb}
\end{eqnarray}
where
\begin{eqnarray}
H^\pm_{C'}=-J\!\!\sum_{\langle i,j\rangle \in C'}\!\!\sigma_i \sigma_j-\!\!\sum_{\begin{subarray}{c}i\in C'\\(i\neq 4')\end{subarray}}\!\!h_{\rm eff}^{\sigma_i}\sigma_i\mp J\sigma_{4'}. \label{Honeycomb'}
\end{eqnarray}
The upper (lower) signs correspond to the case where the value of $\sigma_1$ is fixed to be $+1$ ($-1$). Thus it is only necessary to solve the set of two equations (namely, two six-site problems) given by Eqs.~(\ref{mpm_Honeycomb}) and (\ref{Honeycomb'}) self-consistently for $m^+$ and $m^-$ and then one can calculate the magnetization $m$ from Eq.~(\ref{m}). 

\subsection{\label{2-2}Square lattice}
\begin{figure}[t]
\begin{center}
\includegraphics[width=88mm]{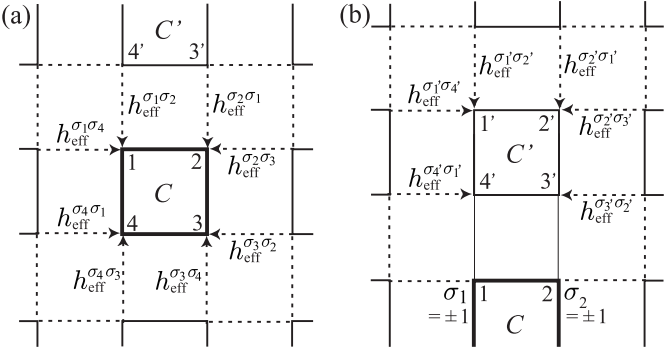}
\caption{\label{fig2}Schematic representation of the square lattice divided into square-shaped clusters. The arrows indicate the effective fields acting on the sites (a) in cluster $C$ and (b) in cluster $C'$ when the spins at site $1$ and site $2$ are fixed to be $+1$ or $-1$, respectively. }
\end{center}
\end{figure}

In the case of the square lattice, there are two bonds between two neighboring clusters as shown in Fig.~\ref{fig2} and this makes matters somewhat more complicated. The spins in cluster $C'$ are strongly affected by the states of the spins at site $1$ and site $2$ of cluster $C$, and thus, including the effect of the feedback, the effective fields acting on site $1$ and site $2$ should depend on the states of themselves. Then, for example, the effective field from site $4'$ to site $1$ can be denoted by $h_{\rm eff}^{\sigma_1\sigma_2}=Jm^{\sigma_1\sigma_2}$, where
\begin{eqnarray}
m^{\sigma_i\sigma_j}=\left\{ \begin{array}{ll}
m^{++} & (\sigma_i=+1,~\sigma_j=+1) \\
m^{+-} & (\sigma_i=+1,~\sigma_j=-1) \\
m^{-+} & (\sigma_i=-1,~\sigma_j=+1) \\
m^{--} & (\sigma_i=-1,~\sigma_j=-1). \label{MF_Square}
\end{array} \right.
\end{eqnarray}
It is clear from the symmetry that the effective field from site $3'$ to site $2$ should be $h_{\rm eff}^{\sigma_2\sigma_1}=Jm^{\sigma_2\sigma_1}$ [see Fig.~\ref{fig2}(a) for the others]. In this way, within our approximation, the states of the four spins in cluster $C$ can be described by the Hamiltonian
\begin{eqnarray}
H_{C}&=&-J\!\!\sum_{\langle i,j\rangle \in C}\!\!\sigma_i \sigma_j-\sum_{i,\bar{i} \in C}h_{\rm eff}^{\sigma_i\sigma_{\bar{i}}}\sigma_i\nonumber\\
&=&-J\left(\sigma_1 \sigma_2+\sigma_2 \sigma_3+\sigma_3 \sigma_4+\sigma_4 \sigma_1\right)\nonumber\\
&&-J\left(m^{\sigma_1\sigma_2}+m^{\sigma_1\sigma_4}\right)\sigma_1\nonumber\\
&&-J\left(m^{\sigma_2\sigma_3}+m^{\sigma_2\sigma_1}\right)\sigma_2\nonumber\\
&&-J\left(m^{\sigma_3\sigma_4}+m^{\sigma_3\sigma_2}\right)\sigma_3\nonumber\\
&&-J\left(m^{\sigma_4\sigma_1}+m^{\sigma_4\sigma_3}\right)\sigma_4, 
 \label{Square}
\end{eqnarray}
where $\sum_{\bar{i} \in C}$ denotes the sum over all nearest neighbors of site $i$ within cluster $C$. The values of the four mean fields shown in Eq.~(\ref{MF_Square}) can be determined by
\begin{eqnarray}
m^{ss'}&=&\langle \sigma_{4'}\rangle \big|_{\sigma_1=s,\sigma_2=s'}\nonumber\\
&=&{\rm Tr}\left(\sigma_{4'} e^{-\beta H^{ss'}_{C'}}\right)/{\rm Tr}\left(e^{-\beta H^{ss'}_{C'}}\right), \label{mpm_Square}
\end{eqnarray}
where
\begin{eqnarray}
H^{ss'}_{C'}&=&-J\!\!\sum_{\langle i,j\rangle \in C'}\!\!\sigma_i \sigma_j-\!\!\!\!\sum_{\begin{subarray}{c}i,\bar{i}\in C'\\\left(\{i,\bar{i}\} \neq \{3',4'\},\{4',3'\}\right)\end{subarray}}\!\!\!\!h_{\rm eff}^{\sigma_i\sigma_{\bar{i}}}\sigma_i\nonumber\\
&&-s' J\sigma_{3'}-s J\sigma_{4'}\nonumber\\
&=&-J\left(\sigma_{1'} \sigma_{2'}+\sigma_{2'} \sigma_{3'}+\sigma_{3'} \sigma_{4'}+\sigma_{4'} \sigma_{1'}\right)\nonumber\\
&&-J\left(m^{\sigma_{1'}\sigma_{2'}}+m^{\sigma_{1'}\sigma_{4'}}\right)\sigma_{1'}\nonumber\\
&&-J\left(m^{\sigma_{2'}\sigma_{3'}}+m^{\sigma_{2'}\sigma_{1'}}\right)\sigma_{2'}\nonumber\\
&&-J\left(m^{\sigma_{3'}\sigma_{2'}}+s'\right)\sigma_{3'}\nonumber\\
&&-J\left(m^{\sigma_{4'}\sigma_{1'}}+s \right)\sigma_{4'}. 
 \label{Square'}
\end{eqnarray}
The notation $H^{ss'}_{C'}$ corresponds to the case where the values of $\sigma_1$ and $\sigma_2$ are fixed to be $s$ and $s'$, respectively ($s,s'=+1$ or $-1$). Solving the set of four equations given by Eqs.~(\ref{mpm_Square}) and (\ref{Square'}) self-consistently, one can obtain the values of the mean fields $m^{++}$, $m^{+-}$, $m^{-+}$, and $m^{--}$. Then the magnetization $m$ [Eq.~(\ref{m})] can be calculated by solving the four-site problem given by Eq.~(\ref{Square}). 

\subsection{\label{2-3}Triangular lattice}
\begin{figure}[t]
\begin{center}
\includegraphics[width=84mm]{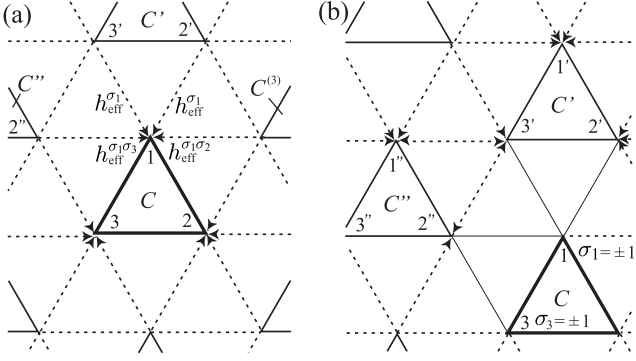}
\caption{\label{fig3}Schematic representation of the triangular lattice divided into triangle-shaped clusters. (a) The effective fields acting on site $1$. (b) Positional relation between three clusters ($C$, $C'$, and $C''$). }
\end{center}
\end{figure}

For the triangular lattice, we perform the calculation based on triangle-shaped three-site clusters shown in Fig.~\ref{fig3} in a basically similar fashion to the previous two cases. 
From the geometry, there are two types of effective fields acting on a site. Let us specifically focus on site $1$ in cluster $C$. There are four bonds from sites in other clusters to site $1$: two of them are from cluster $C'$ and the other two are from cluster $C''$ and $C^{(3)}$, respectively. 

Since cluster $C'$ is connected to cluster $C$ at only one point, the spins at site $2'$ and site $3'$ are {\it directly} affected only by the state of the spin at site $1$. Thus we assume that the two effective fields from sites $2'$ and $3'$ depends only on the state of the spin at site $1$: $2h_{\rm eff}^{\sigma_1}=2Jm^{\sigma_1}$ [see, Eq.~(\ref{MF_Honeycomb}) for the definition of $m^{\sigma_i}$]. 

As for cluster $C''$ ($C^{(3)}$), on the other hand, there are connections with cluster $C$ at two points, site $1$ and site $3$ (site $1$ and site $2$). Thus, in our procedure the effective field from cluster $C''$ ($C^{(3)}$) to site $1$ should depend on the spins at these two sites: $h_{\rm eff}^{\sigma_1\sigma_3}=Jm^{\sigma_1\sigma_3}$ ($h_{\rm eff}^{\sigma_1\sigma_2}=Jm^{\sigma_1\sigma_2}$) [see, Eq.~(\ref{MF_Square}) for the definition of $m^{\sigma_i\sigma_j}$]. When taken together, the total effective field acting on site $1$ is given by $2h_{\rm eff}^{\sigma_1}+h_{\rm eff}^{\sigma_1\sigma_3}+h_{\rm eff}^{\sigma_1\sigma_2}$ [see Fig.~\ref{fig3}(a)]. 

Eventually, for the case of the triangular lattice, we obtain the following Hamiltonian describing the three spins in cluster $C$: 
\begin{eqnarray}
H_{C}&=&-J\!\!\sum_{\langle i,j\rangle \in C}\!\!\sigma_i \sigma_j-\sum_{i \in C}2h_{\rm eff}^{\sigma_i}\sigma_i-\sum_{i,\bar{i} \in C}h_{\rm eff}^{\sigma_i\sigma_{\bar{i}}}\sigma_i\nonumber\\
&=&-J\left(\sigma_{1} \sigma_{2}+\sigma_{2} \sigma_{3}+\sigma_{3} \sigma_{1}\right)\nonumber\\
&&-J\left(2m^{\sigma_{1}}+m^{\sigma_{1}\sigma_{2}}+m^{\sigma_{1}\sigma_{3}}\right)\sigma_{1}\nonumber\\
&&-J\left(2m^{\sigma_{2}}+m^{\sigma_{2}\sigma_{3}}+m^{\sigma_{2}\sigma_{1}}\right)\sigma_{2}\nonumber\\
&&-J\left(2m^{\sigma_{3}}+m^{\sigma_{3}\sigma_{1}}+m^{\sigma_{3}\sigma_{2}}\right)\sigma_{3}. \label{Triangular_Hamiltonian}
\end{eqnarray}
The magnetization $m$ can be calculated from this Hamiltonian via Eq.~(\ref{m}). 

In a similar way to the previous two cases, let us determine the values of the mean fields $m^{+}$, $m^{-}$, $m^{++}$, $m^{+-}$, $m^{-+}$, and $m^{--}$. 
First, fixing the value of $\sigma_1$ and considering the three-site problem of cluster $C'$, one can obtain the equations for the mean fields $m^+$ and $m^-$, 
\begin{eqnarray}
m^\pm\!=\!\langle \sigma_{2'}\rangle \big|_{\sigma_1=\pm 1}\!=\!{\rm Tr}\left(\sigma_{2'} e^{-\beta H^\pm_{C'}}\right)/{\rm Tr}\left(e^{-\beta H^\pm_{C'}}\right), \label{mpm_Triangular1}
\end{eqnarray}
where
\begin{eqnarray}
H^{\pm}_{C'}&=&-J\!\!\sum_{\langle i,j\rangle \in C'}\!\!\sigma_i \sigma_j-\sum_{i \in C'}2h_{\rm eff}^{\sigma_i}\sigma_i\nonumber\\
&&-\!\!\!\!\sum_{\begin{subarray}{c}i,\bar{i}\in C'\\\left(\{i,\bar{i}\} \neq \{2',3'\},\{3',2'\}\right)\end{subarray}}\!\!\!\!h_{\rm eff}^{\sigma_i\sigma_{\bar{i}}}\sigma_i\mp J\left(\sigma_{2'}+\sigma_{3'}\right)\nonumber\\
&=&-J\left(\sigma_{1'} \sigma_{2'}+\sigma_{2'} \sigma_{3'}+\sigma_{3'} \sigma_{1'}\right)\nonumber\\
&&-J\left(2m^{\sigma_{1'}}+m^{\sigma_{1'}\sigma_{2'}}+m^{\sigma_{1'}\sigma_{3'}}\right)\sigma_{1'}\nonumber\\
&&-J\left(2m^{\sigma_{2'}}+m^{\sigma_{2'}\sigma_{1'}}\pm 1\right)\sigma_{2'}\nonumber\\
&&-J\left(2m^{\sigma_{3'}}+m^{\sigma_{3'}\sigma_{1'}}\pm 1\right)\sigma_{3'}.
 \label{Triangular'}
\end{eqnarray}
The upper (lower) signs correspond to the case where the value of $\sigma_1$ is fixed to be $+1$ ($-1$). Due to the symmetry, of course, one can obtain the same value of $m^\pm$ by calculating $\langle \sigma_{3'}\rangle \big|_{\sigma_1=\pm 1}$ instead of Eq.~(\ref{mpm_Triangular1}). 

Next, we calculate the other type of the mean fields: $m^{++}$, $m^{+-}$, $m^{-+}$, and $m^{--}$. To this end we now focus on, for example, cluster $C''$ [see Fig.~\ref{fig3}(b)]. In a similar manner to the case of the square lattice, we can derive the following equations for the four mean fields: 
\begin{eqnarray}
m^{ss'}&=&\langle \sigma_{2''}\rangle \big|_{\sigma_1=s,\sigma_3=s'}\nonumber\\
&=&{\rm Tr}\left(\sigma_{2''} e^{-\beta H^{ss'}_{C''}}\right)/{\rm Tr}\left(e^{-\beta H^{ss'}_{C''}}\right), \label{mpm_Triangular2}
\end{eqnarray}
where
\begin{eqnarray}
H^{ss'}_{C''}&=&-J\!\!\sum_{\langle i,j\rangle \in C''}\!\!\sigma_i \sigma_j-\sum_{\begin{subarray}{c}i\in C''\\(i\neq 2'')\end{subarray}}2h_{\rm eff}^{\sigma_i}\sigma_i\nonumber\\
&&-\!\!\sum_{i,\bar{i}\in C''}\!\!h_{\rm eff}^{\sigma_i\sigma_{\bar{i}}}\sigma_i-J\left(s+s'\right)\sigma_{2''}\nonumber\\
&=&-J\left(\sigma_{1''} \sigma_{2''}+\sigma_{2''} \sigma_{3''}+\sigma_{3''} \sigma_{1''}\right)\nonumber\\
&&-J\left(2m^{\sigma_{1''}}+m^{\sigma_{1''}\sigma_{2''}}+m^{\sigma_{1''}\sigma_{3''}}\right)\sigma_{1''}\nonumber\\
&&-J\left(m^{\sigma_{2''}\sigma_{3''}}+m^{\sigma_{2''}\sigma_{1''}}+s+s'\right)\sigma_{2''}\nonumber\\
&&-J\left(2m^{\sigma_{3''}}+m^{\sigma_{3''}\sigma_{1''}}+m^{\sigma_{3''}\sigma_{2''}}\right)\sigma_{3''}. 
 \label{Triangular'2}
\end{eqnarray}

In the case of the triangular lattice, one have to solve the set of six nonlinear equations in terms of the six parameters $m^+$, $m^-$, $m^{++}$, $m^{+-}$, $m^{-+}$, and $m^{--}$, which are given by Eqs.~(\ref{mpm_Triangular1})-(\ref{Triangular'2}). Nonetheless, since the cluster we selected here consists of only three sites, it is easy to solve the set of equations. 

\begin{table*}[t]
\caption{\label{table1}Reduced critical temperatures ($k_BT_c/J$) for the Ising model on various types of lattices from different approximations and exact or series values. }
\begin{ruledtabular}
\begin{tabular}{lccccccc}
~~Lattice & MFT($z$) & Exact or series & BPW & ORF & SCCF & SMF & CCMF~~ \\ \hline 
~~Honeycomb & 3& 1.519 & 1.820 & - & - & 1.464 & 1.593~~ \\
~~Square & 4& 2.269 & 2.885 & - & 2.595 & 2.142 & 2.362~~ \\
~~Triangular & 6& 3.641 & 4.933 & - & 4.788 & 2.628 (3.543\footnote{From the corrected SMF with $\alpha=0.8$ (see Ref.~\onlinecite{Zhuravlev}). }) & 4.519~~ \\
~~Simple cubic & 6& 4.510 & 4.933 & 3.955 & 4.788 & 4.570 & 4.753~~ 
\end{tabular}
\end{ruledtabular}
\end{table*}
\begin{table*}[t]
\caption{\label{table2}Comparison of the results obtained by the CCMF method for the square-lattice Ising model with those obtained by the BPW method, its extended version proposed by Etxebarria $et$ $al.$ (Ref.~\onlinecite{Etxebarria}), Kikuchi's square approximation (Refs.~\onlinecite{Kikuchi} and \onlinecite{Morita}), and exact values. $C_{1}(T_c)$ and $C_{2}(T_c)$ are the values of the correlation functions between first and second neighbors, respectively, at $T_c$. $N_c$ is the size of the cluster, and $M$ is the number of nonlinear equations involved.}
\begin{ruledtabular}
\begin{tabular}{lccccc}
~~~~Method & $N_c$ & $M$ & $k_BT_c/J$ & $C_{1}(T_c)$ & $C_{2}(T_c)$ ~~~~ \\ \hline 
~~~~Exact & & & 2.269 & 0.707 & 0.637~~~~  \\
~~~~BPW & 5& 1 & 2.885 & 0.333 & 0.111~~~~  \\
~~~~Etxebarria $et$ $al.$ & 12 & 4 & 2.351 & 0.607 & 0.501~~~~ \\
~~~~Kikuchi & 4 & 3 & 2.426 & 0.562 & 0.438~~~~ \\
~~~~CCMF & 4 & 4 & 2.362 & 0.608 & 0.495~~~~ 
\end{tabular}
\end{ruledtabular}
\end{table*}

\subsection{\label{2-4}Simple cubic lattice}
\begin{figure}[t]
\begin{center}
\includegraphics[height=30mm]{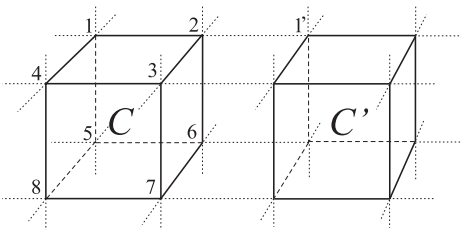}
\caption{\label{fig4}Schematic representation of eight-site clusters for the simple cubic lattice. }
\end{center}
\end{figure}
As an example of three-dimensional (3D) lattices, we show the application of our method to the case of the simple cubic lattice, where the calculation is performed based on eight-site cubic clusters. In this case, clusters are connected by four points of each face (see Fig.~\ref{fig4}), and thus we have $2^4$ mean fields ($m^{++++},m^{+++-},\dots,m^{----}$). For example, the effective field from site $1'$ to site $2$ is defined by a function of the states of the four spins $\sigma_2$, $\sigma_3$, $\sigma_6$, and $\sigma_7$. Due to symmetries, only 12 of the $2^4$ mean fields are actually independent. The calculation is straightforward extension of that in the case of the square lattice, so we display only the final result in Sec.~\ref{3}. 

\section{\label{3}Discussion of results for the Ising model}
Before presenting the results obtained by the above procedures, we will briefly review the screened magnetic field (SMF) approximation, which was recently proposed by Zhuravlev.\cite{Zhuravlev} In his method, one considers an effective one-site problem of a certain spin (the central spin) in an effective magnetic field like the standard MFT. The influence of the interactions between the central spin and all others should decrease quickly with the distance between them. In the SMF, he took into account this ``screening effect'' in the following way. Introducing the factor $e^{-r^2/a_0^2}$, he assumed that the effective field acting on the central site can be represented by
\begin{eqnarray}
h_{\rm eff}=Jm\sum_{\{\sigma\}}e^{-r_\sigma^2/a_0^2}, \label{Zhuravlev}
\end{eqnarray}
where the sum is over all spins except for the central spin itself. As a characteristic length $a_0$, the lattice constant was chosen in Ref.~\onlinecite{Zhuravlev}. For the calculation of the critical points, one have only to replace the effective field in the MFT by Eq.~(\ref{Zhuravlev}). Then one obtains
\begin{eqnarray}
k_{B}T_c/J=\sum_{\{\sigma\}}e^{-r_\sigma^2/a_0^2}. \label{TcSMF}
\end{eqnarray}

In Table~\ref{table1}, we summarize the results of the critical temperature $T_c$ obtained by the CCMF and other methods. We can see that the results obtained by our CCMF method are very close to the exact results or approximate value from the series-expansion method.\cite{Onsager,Fisher2,Wu} Particularly, for the honeycomb and square lattices, our results ($k_BT_c/J=1.593$ and $2.362$) overestimate the corresponding exact values by only less than $5$ \%. For the triangular lattice, the accuracy is inferior as compared with the above two cases, but is better than those of the other methods [except for the ``corrected'' SMF (Ref.~\onlinecite{Zhuravlev}) with the correction coefficient $\alpha=0.8$]. 

In most previous approaches, such as MFT, BPW, and SCCF, the results are dependent only on the coordination number $z$. In contrast, our cluster approach can distinguish between two lattices with different geometries but equal coordination numbers, as well as the SMF method. For example, the obtained critical points of the triangular and simple-cubic lattices are different, although they have the same coordination number $z=6$. This is because we selected different clusters depending on the geometries of the lattices. For the 3D simple cubic lattice, unfortunately, our result of the critical point represents only a little improvement from that of the SCCF approximation. This is attributed to the fact that the importance of including cluster correlations is relatively low compared with 2D cases, since mean-field approaches are generally speaking expected to work better in higher dimensions. 

Next, focusing on the case of the square lattice, let us compare the results obtained by the CCMF method with those obtained by the extended version of the BPW method proposed by Etxebarria $et$ $al$.~\cite{Etxebarria} and Kikuchi's square approximation.\cite{Kikuchi,Morita} In the method of Etxebarria $et$ $al$., a relatively large cluster (consisting of 12 spins) is considered, and one introduces one effective field and three effective couplings between the ``boundary'' spins, which are determined by the condition that the periodicity of the system is preserved (for detail, see Ref.~\onlinecite{Etxebarria}). The so-called cluster-variation method (CVM) proposed by Kikuchi~\cite{Kikuchi} provides a systematic hierarchy of approximations, each of which is specified by the set of ``basic clusters'' used for the calculation. For example, in the ``square'' approximation~\cite{Kikuchi,Morita} of the CVM, the largest basic cluster is a $2\times 2$ square. 

In Table~\ref{table2}, we summarize the results of the critical temperature $T_c$ and values of the connected correlation functions between first and second neighbors at $T_c$, obtained by four different cluster-based approaches. In the CCMF calculation, the correlation functions between site $i$ and site $j$ can be obtained by $\langle \sigma_i\sigma_j\rangle-\langle \sigma_i\rangle^2={\rm Tr}\left(\sigma_i\sigma_j \exp [-\beta H_C]\right)/{\rm Tr}\left(\exp [-\beta H_C]\right)-m^2~(i,j\in C)$. The extended BPW approximation gives the most accurate value ($k_BT_c/J=2.351$) compared with the other approximation methods in Tables~\ref{table1} and \ref{table2}, although one have to deal with a 12-site problem. Kikuchi's square approximation gives $k_BT_c/J=2.426$. The CVM results are usually more and more accurate as the size of the considered basic clusters increases. Using the Tanoji approximation,~\cite{Fujiki} where a nine-site cluster is used as the largest basic cluster and $44$ variational parameters ($M=44$) are needed, one can obtain the more accurate result ($k_BT_c/J=2.346$). Our CCMF procedure can achieve the same level of accuracy by treating only four sites. (However, note that in the CCMF approach it is necessary to solve a set of four slightly different four-site problems to obtain the values of all unknown mean fields.)

\section{\label{4}Cluster-size dependence and critical exponents}
In Sec.~\ref{2}, we have demonstrated the application of the CCMF method to the Ising model by using a cluster of appropriate size and shape for each lattice type. For the square lattice, we employed the clusters illustrated in Fig.~\ref{fig2}. Hereafter we refer to this procedure as the square approximation of the CCMF method for descriptive purposes. 
\begin{table}[t]
\caption{\label{table3} The values of $k_BT_c/J$ and the mean-field critical coefficients for the square-lattice Ising model in the single site (SCCF), pair, and square approximations of the CCMF method. }

\begin{ruledtabular}
\begin{tabular}{lccccccc}
~Approximation & $k_BT_c/J$ & $\bar{m}_s$ & $J\bar{\chi}_+$ & $J\bar{\chi}_-$ \\ \hline 
~SCCF & 2.595225 & 2.534519 & 0.716621 & 0.358311~ \\
~Pair               & 2.437857 & 2.922637 & 1.130168 & 0.565084~ \\
~Square             & 2.362099 & 3.551663 & 1.537636 & 0.768818~  
\end{tabular}
\end{ruledtabular}
\end{table}
\begin{table}[t]
\caption{\label{table4} Comparison of the CAM estimates and exact values of critical exponents.}

\begin{ruledtabular}
\begin{tabular}{lccccccc}
 & $k_BT_c/J$ & $\beta$ & $\gamma=\gamma '$  \\ \hline 
~~CAM            &  & 0.343 & 2.138~~ \\
~~($k_BT^*_c/J$) &  & (2.331) & (2.118)~~ \\ \hline
~~Exact     & 2.269 & 0.125 & 1.75 ~~ 
\end{tabular}
\end{ruledtabular}
\end{table}

In Table~\ref{table3} we show the values of $k_BT_c/J$ for the square-lattice Ising model in three different approximations: the single site (i.e., SCCF), pair, and square approximations of the CCMF method. In the ``pair'' approximation, we adopt a nearest-neighbor pair of spins as a basic cluster for the calculation (for detail, see the Appendix). It can be seen that the value of the critical temperature becomes more and more close to the exact one ($k_BT_c/J=2.269185\cdots$) with increasing the size of the used cluster. However, in all the cases, the obtained critical exponents are classical ones, as in most other mean-field approaches. Namely, the temperature dependence of the spontaneous magnetization just below $T_c$ is given by
\begin{eqnarray}
m=\bar{m}\left| \frac{T-T_c}{T_c}\right| ^{1/2}~~{\rm for}~~T\lesssim T_c,  \label{spontaneous_magnetization}
\end{eqnarray}
and the zero-field susceptibility just above $T_c$ and just below $T_c$ are given by
\begin{eqnarray}
\chi=\left\{\begin{array}{lcc}
\displaystyle \bar{\chi}_+\left(\frac{T-T_c}{T_c}\right)^{-1}& {\rm for} & T\gtrsim T_c \\
\displaystyle \bar{\chi}_-\left|\frac{T-T_c}{T_c}\right|^{-1}& {\rm for} & T\lesssim T_c. \end{array} \right.  \label{zero-field_susceptibility}
\end{eqnarray}
The values of the critical coefficients $\bar{m}$ and $\bar{\chi}_\pm$ are also shown in Table~\ref{table3}. 

The coherent-anomaly method (CAM) proposed by Suzuki~\cite{Suzuki} allows one to estimate non-classical critical exponents from a series of mean-field-type approximations. Now, assuming that the approximations in the CCMF scheme constitute a part of {\it a canonical series},\cite{Suzuki} let us estimate the non-classical critical exponents for the square-lattice Ising model. According to the CAM, the critical coefficients $\bar{m}$ and $\bar{\chi}_\pm$ should behave asymptotically as 
\begin{subequations}
\label{scaling}
\begin{eqnarray}
\bar{m}&\sim& c_1 \delta(T_c)^{-(1/2-\beta)},\\
\bar{\chi}_+&\sim& c_2 \delta(T_c)^{-(\gamma-1)},\\
\bar{\chi}_-&\sim& c_3 \delta(T_c)^{-(\gamma '-1)},
\end{eqnarray}
\end{subequations}
for
\begin{eqnarray}
\delta=\frac{T_c-T_c^*}{T_c^*}\rightarrow 0, 
\end{eqnarray}
where $\{c_i \}$ are certain constants, $T_c^*$ denotes the exact value of the critical temperature, and $T_c$ is the approximate critical temperature obtained in each approximation. Since each equation has three unknown variables, $c_i$, $T_c^*$, and the critical exponent ($\beta$, $\gamma$, or $\gamma '$), we need the approximate values of $T_c$ and the corresponding critical coefficients ($\bar{m}$, $\bar{\chi}_+$, or $\bar{\chi}_-$) for, at least, three different levels of approximations. 

In Table~\ref{table4}, we show the values of $\beta$, $\gamma$, and $\gamma '$ obtained from Eq.~(\ref{scaling}) and the data obtained from the three levels of approximations of the CCMF method presented in Table~\ref{table3}. Although non-classical exponents are obtained, the values of them are not very accurate. This may be because the size of clusters used in the approximations is relatively small.~\cite{Katori} It is expected that more accurate results are obtained by taking a larger cluster like that used in the Tanoji approximation of the CVM.~\cite{Fujiki} However, since the initial lattice symmetry is artificially broken,\cite{Etxebarria,Galam} the CCMF theory is not directly applicable for a cluster larger than a certain critical size (e.g., the $2\times 2$ cluster for the square lattice) in the present form. 
 
\section{\label{5}Application to the Heisenberg model}
In this section, we consider the application of the CCMF theory to a spin-$\frac{1}{2}$ Heisenberg ferromagnet in a uniform magnetic field as an example of quantum spin systems. The Hamiltonian of the system is given by 
\begin{eqnarray}
H&=&-J\sum_{\langle i,j\rangle }\mathbf{S}_{i}\cdot\mathbf{S}_{j}-h\sum_{i}S^z_i\nonumber\\
&=&-J\sum_{\langle i,j\rangle }\left[\frac{1}{2}\left(S_i^+S_j^-+S_i^-S_j^+\right)+S_i^zS_j^z\right]\nonumber\\
&&-h\sum_{i}S^z_i. \label{Heisenberg}
\end{eqnarray}
Here $\mathbf{S}_{i}=(S_i^x,S_i^y,S_i^z)$ is the usual spin operator at site $i$, which satisfies the commutation relation $[S_i^\mu,S_j^\nu]=i\epsilon_{\mu\nu\lambda} S_i^\lambda\delta_{ij}$, $S_i^{\pm}=S_i^x \pm iS_i^y$ are the spin raising and lowering operators, and $h$ denotes an applied magnetic field. 

One of the advantage of cluster-based approaches, such as the BPW and CCMF methods, is that the effects of the spin-flip term, $S_i^+S_j^-+S_i^-S_j^+$, can be taken into account by the straightforward application. Now, we focus again on the case of the square lattice, and calculate the magnetization of the system by using the CCMF approach. In the square approximation (see Fig.~\ref{fig2}), we first introduce the four effective fields $h_{\rm eff}^{\sigma_i\sigma_j}=Jm^{\sigma_i\sigma_j}$, where
\begin{eqnarray}
m^{\sigma_i\sigma_j}&=&m^{++}\left(\frac{1}{2}+S^z_i\right)\left(\frac{1}{2}+S^z_j\right)\nonumber\\
&&+m^{+-}\left(\frac{1}{2}+S^z_i\right)\left(\frac{1}{2}-S^z_j\right)\nonumber\\
&&+m^{-+}\left(\frac{1}{2}-S^z_i\right)\left(\frac{1}{2}+S^z_j\right)\nonumber\\
&&+m^{--}\left(\frac{1}{2}-S^z_i\right)\left(\frac{1}{2}-S^z_j\right)\nonumber\\
&=&\left\{ \begin{array}{ll}
m^{++} & (\sigma_i=\uparrow,~\sigma_j=\uparrow), \\
m^{+-} & (\sigma_i=\uparrow,~\sigma_j=\downarrow), \\
m^{-+} & (\sigma_i=\downarrow,~\sigma_j=\uparrow), \\
m^{--} & (\sigma_i=\downarrow,~\sigma_j=\downarrow). 
\end{array} \right.
\end{eqnarray}
Here, $\sigma_i=\uparrow,\downarrow$ represents the $z$ component of the spin at site $i$. The procedure for the calculations of these mean fields and the magnetization $m$ is almost the same as that in Sec.~\ref{2-2}. The difference is only that the Ising interaction between neighboring spins in a cluster is replaced by the Heisenberg one, $\mathbf{S}_{i}\cdot\mathbf{S}_{j}$, in this case. 

\begin{figure}[t]
\begin{center}
\includegraphics[height=55mm]{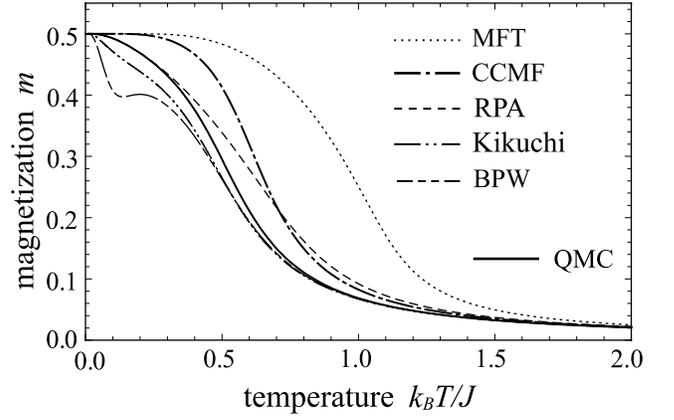}
\caption{\label{fig5}The temperature dependence of the magnetization of an isotropic Heisenberg ferromagnet at $h/J=0.1$. Comparison of the result obtained from the CCMF approach with those of the MFT, BPW, RPA, Kikuchi's square approximation, and the QMC calculations. The error bar of the QMC results are smaller than the line width. }
\end{center}
\end{figure}
The temperature dependence of the magnetization $m$ obtained by the CCMF approach is compared with those from the MFT, BPW, the so-called random-phase approximation (RPA),\cite{Tyablikov,Frobrich} Kikuchi's square approximation,\cite{Morita} and quantum Monte Carlo (QMC) calculations~\cite{Henelius,Yamamoto,Todo,Albuquerque} in Fig.~\ref{fig5}. The BPW approximation, which was first applied to the isotropic Heisenberg model by Weiss in Ref.~\onlinecite{Weiss}, gives a qualitatively incorrect behavior in the low-temperature region. This behavior becomes more pronounced at lower magnetic field strengths. Therefore, when applied to other systems, the BPW method has a risk that one cannot decide whether the behavior of the obtained results is an essential property of the system or an artifact of the approximation. The similar behavior in the low temperature region is also observed in the result of Kikuchi's approximation. In the ``Bethe-type'' or ``effective-field'' approaches, such as the BPW method~\cite{Bethe,Peierls,Weiss} and its extended versions,\cite{Etxebarria,Minami} the values of the effective fields (and effective couplings) are determined by the periodic boundary conditions and thus these approaches generally overestimate the correlation effects between the spins.

The Green's function method with the RPA decoupling approximation (or ``Tyablikov decoupling'' method) is one of the most popular approximation schemes to deal with quantum spin systems. As is shown in the figure, the RPA is a good approximation over a wide temperature range. However, since the treatment for the Ising term of the Hamiltonian is identical to that of the MFT, the approximation should get worse as the Ising-type anisotropy of the system increases. Especially, when the RPA scheme is applied to the Ising model, the obtained results are the same as that of the MFT. Besides, a special attention is required in applying the RPA scheme to a relatively complicated system as is pointed out in our recent study.\cite{Yamamoto} 

In the ``Weiss-type'' approaches, the effective fields are defined in terms of some kind of mean fields (e.g., $h_{\rm eff}=zJm$ in the simple MFT), and those values are determined by the corresponding self-consistency conditions. Oguchi's method, the SCCF, and CCMF approaches are categorized into this group. These approaches generally underestimate the correlation effects between the spins and therefore overestimate the magnetization. The CCMF method also overestimates the magnetization as is shown in Fig.~\ref{fig5}, but the difference from the QMC data is much smaller than that of the MFT. Moreover, above a certain temperature ($k_BT/J\sim 0.7$), the accuracy exceeds that of the RPA scheme. The results obtained here and in Sec.~\ref{3} indicate that our CCMF method can give qualitatively (and even quantitatively) correct results over a wide range of temperature and strength of the exchange anisotropy. 

\section{\label{6}SUMMARY}
In this paper, we have proposed a cluster-based mean-field approach, which we refer to as the CCMF theory and have demonstrated its applications to the Ising and isotropic Heisenberg models as examples. By using the CCMF approach, one can take into account the effects of spin correlations between clusters beyond the standard mean-field level. Since a cluster of different size and shape is used for each lattice type, the obtained results are dependent not only on the coordination number $z$ but also on the geometry of the lattice. 

The results obtained from the CCMF method are in good agreement with the corresponding exact values, series expansion, or QMC data for the both Ising and isotropic Heisenberg models. Especially, for the Ising model on the honeycomb and square lattices, the calculated results of the critical temperature are very close (overestimated by only a few percent) to the exact ones. 

As well as being very accurate, the advantage of this method is that it is widely applicable. Since the calculation in the CCMF approach is based on clusters consisting of several spins, in addition to the Ising-type interaction, the contributions of other types of interactions (e.g., the Heisenberg-type exchange interaction as we demonstrated) can be taken into account in a straightforward way. Although we have considered here only the two simple cases, this method is expected to be useful also for studies of more complicated and interesting systems. For example, the CCMF method should work well for the {\it XXZ} model (Heisenberg-Ising model) with easy-plane anisotropy~\cite{Yamamoto,Hu} and, of course, with Ising-type anisotropy, which lies between the two models we considered here. Also, the extensions to systems with four-spin (ring) exchange interactions,\cite{Buzano} higher spins, and random systems~\cite{Soukoulis,Grest} should be considered in future studies.

\acknowledgments
The author would like to thank S. Kurihara for valuable comments and discussions. This work was supported by a Grant-in-Aid from the Japan Society for the Promotion of Science. 
\begin{appendix}
\section{Pair Approximation of CCMF}\label{appA}

\begin{figure}[t]
\begin{center}
\includegraphics[height=40mm]{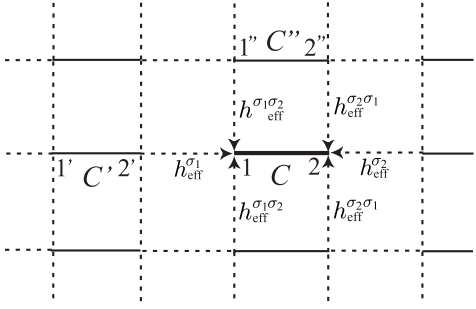}
\caption{\label{fig6}Schematic representation of the square lattice divided into pairs of neighboring sites. The arrows indicate the effective fields acting on the sites in cluster $C$.  }
\end{center}
\end{figure}

In the pair approximation for the square-lattice Ising model, the states of a nearest-neighbor pair of spins (see Fig.~\ref{fig6}) are described by the Hamiltonian 
\begin{eqnarray}
H_{C}&=&-J\!\!\sum_{\langle i,j\rangle \in C}\!\!\sigma_i \sigma_j-\sum_{i \in C}h_{\rm eff}^{\sigma_i}\sigma_i-\sum_{i,\bar{i} \in C}2h_{\rm eff}^{\sigma_i\sigma_{\bar{i}}}\sigma_i\nonumber\\
&=&-J\sigma_{1} \sigma_{2}-J\left(m^{\sigma_{1}}+2m^{\sigma_{1}\sigma_{2}}\right)\sigma_{1}\nonumber\\
&&-J\left(m^{\sigma_{2}}+2m^{\sigma_{2}\sigma_{1}}\right)\sigma_{2}. \label{Pair_Hamiltonian}
\end{eqnarray}
Oguchi's pair approximation~\cite{Oguchi} is obtained by requiring the equality $m^\sigma=m^{\sigma \sigma '}=m$, where $m=(1/2)\sum_C\langle \sigma_i\rangle$, in the above Hamiltonian. In the CCMF method, the values of the mean fields are determined by solving the set of self-consistent equations
\begin{eqnarray}
m^\pm\!=\!\langle \sigma_{2'}\rangle \big|_{\sigma_1=\pm 1}\!=\!{\rm Tr}\left(\sigma_{2'} e^{-\beta H^\pm_{C'}}\right)/{\rm Tr}\left(e^{-\beta H^\pm_{C'}}\right) \label{mpm_Pair1}
\end{eqnarray}
and
\begin{eqnarray}
m^{ss'}&=&\langle \sigma_{1''}\rangle \big|_{\sigma_1=s,\sigma_2=s'}\nonumber\\
&=&{\rm Tr}\left(\sigma_{1''} e^{-\beta H^{ss'}_{C''}}\right)/{\rm Tr}\left(e^{-\beta H^{ss'}_{C''}}\right), \label{mpm_Pair2}
\end{eqnarray}
where
\begin{eqnarray}
H^{\pm}_{C'}&=&-J\!\!\sum_{\langle i,j\rangle \in C'}\!\!\sigma_i \sigma_j-h_{\rm eff}^{\sigma_{1'}}\sigma_{1'}-\sum_{i,\bar{i} \in C}2h_{\rm eff}^{\sigma_i\sigma_{\bar{i}}}\sigma_i \mp J \sigma_{2'} \nonumber\\
&=&-J \sigma_{1'} \sigma_{2'}-J\left(m^{\sigma_{1'}}+2m^{\sigma_{1'}\sigma_{2'}}\right)\sigma_{1'}\nonumber\\
&&-J\left(2m^{\sigma_{2'}\sigma_{1'}}\pm 1\right)\sigma_{2'}
 \label{Pair'}
\end{eqnarray}
and
\begin{eqnarray}
H^{ss'}_{C''}&=&-J\!\!\sum_{\langle i,j\rangle \in C''}\!\!\sigma_i \sigma_j-\sum_{i\in C''}h_{\rm eff}^{\sigma_i}\sigma_i -\sum_{i,\bar{i} \in C}h_{\rm eff}^{\sigma_i\sigma_{\bar{i}}}\sigma_i\nonumber\\
&&-sJ\sigma_{1''}-s'J\sigma_{2''}\nonumber\\
&=&-J\sigma_{1''} \sigma_{2''}-J\left(m^{\sigma_{1''}}+m^{\sigma_{1''}\sigma_{2''}}+s\right)\sigma_{1''}\nonumber\\
&&-J\left(m^{\sigma_{2''}}+m^{\sigma_{2''}\sigma_{1''}}+s'\right)\sigma_{2''}. 
 \label{Pair'2}
\end{eqnarray}
Then the magnetization $m$ is calculated from Eqs.~(\ref{m}) and~(\ref{Pair_Hamiltonian}). 
\end{appendix}

%

\end{document}